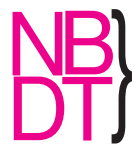

**ORIGINAL ARTICLE**

**Neural data science / analysis**

# The neural dynamics underlying prioritisation of task-relevant information

Tijl Grootswagers[1,2] | Amanda K. Robinson[2] | Sophia M. Shatek[2] | Thomas A. Carlson[2]

[1]The MARCS Institute for Brain, Behaviour and Development, Western Sydney University, Sydney, Australia

[2]School of Psychology, University of Sydney, Australia

**Correspondence**
Tijl Grootswagers, Western Sydney University, Sydney, NSW, Australia
Email:
t.grootswagers@westernsydney.edu.au

**Funding information**
T.A.C: Australian Research Council Discovery Projects (DP160101300 and DP200101787); A.K.R: Australian Research Council Discovery Early Career Research Award (DE200101159)

The human brain prioritises relevant sensory information to perform different tasks. Enhancement of task-relevant information requires flexible allocation of attentional resources, but it is still a mystery how this is operationalised in the brain. We investigated how attentional mechanisms operate in situations where multiple stimuli are presented in the same location and at the same time. In two experiments, participants performed a challenging two-back task on different types of visual stimuli that were presented simultaneously and superimposed over each other. Using electroencephalography and multivariate decoding, we analysed the effect of attention on the neural responses to each individual stimulus. Whole brain neural responses contained considerable information about both the attended and unattended stimuli, even though they were presented simultaneously and represented in overlapping receptive fields. As expected, attention increased the decodability of stimulus-related information contained in the neural responses, but this effect was evident earlier for stimuli that were presented at smaller sizes. Our results show that early neural responses to stimuli in fast-changing displays contain remarkable in-formation about the sensory environment but are also modulated by attention in a manner dependent on perceptual characteristics of the relevant stimuli. Stimuli, code, and data for this study can be found at `https://osf.io/7zhwp/`.

# 1 | INTRODUCTION

To efficiently perform a task, our brains continuously prioritise and select relevant information from a constant stream of sensory input. All sensory input is automatically and unconsciously processed, but the depth of processing varies depending on the task and input characteristics (Grootswagers et al., 2019a; King et al., 2016; Mohsenzadeh et al., 2018; Robinson et al., 2019; Rossion et al., 2015; Rousselet et al., 2002). At what stage in the response is task-relevant information prioritised? Neurophysiological methods such as electroencephalography (EEG) and magnetoencephalography (MEG) have offered insight into the time-scales at which selective attention operates in the human brain (Goddard et al., 2019). For example, a stimulus that is presented in an attended location evokes a stronger neural response around 100ms (e.g., Mangun, 1995; Mangun et al., 1993). Similarly, when a certain feature of a stimulus is attended, the neural correlate of this feature is enhanced (Martinez-Trujillo and Treue, 2004; Maunsell and Treue, 2006), with effects for basic features (e.g., colour) starting as early as 100ms (e.g., Zhang and Luck, 2009). In a sequence of stimuli, temporal selection of task-relevant target stimuli is reported around 270ms (Kranczioch et al., 2005, 2003; Marti and Dehaene, 2017; Sergent et al., 2005; Tang et al., 2020; Wyart et al., 2015). A question that has received considerably less focus is how these mechanisms interact in situations where multiple stimuli are presented in the same location and at the same time. Determining the stages of processing affected by attention in these situations is important for understanding selective attention as a whole, and for constructing an overarching theory of attention.

Studying neural responses to simultaneously presented stimuli is difficult, as the stimulus-specific signals are overlapping. One solution is to display stimuli at different presentation rates and analyse neural responses in the matching frequency bands (e.g., Ding et al., 2006; Müller et al., 2006), but this approach does not allow studying the underlying temporal dynamics. Another approach is to use multivariate decoding methods, which have recently provided new opportunities to study attentional effects on information at the individual stimulus level (e.g., Alilović et al., 2019; Goddard et al., 2019; Marti and Dehaene, 2017; Smout et al., 2019). These methods also allow to decode the overlapping neural signals evoked by stimuli presented close in time (e.g., Grootswagers et al., 2019a; Marti and Dehaene, 2017; Robinson et al., 2019), even when these stimuli are not task-relevant (Grootswagers et al., 2019b; Marti and Dehaene, 2017; Robinson et al., 2019). Multivariate decoding methods can therefore be used to disentangle information from simultaneously presented stimuli and investigate the temporal dynamics of attentional mechanisms operating on the stimuli.

We conducted two experiments to investigate the effect of attention on the representations of simultaneously presented objects and letters. Participants were shown images of objects overlaid with alphabet letters, or vice versa, in rapid succession and performed a cognitively demanding 2-back task on either the object or the letters, which required attending to one of the two simultaneously presented stimuli. We then performed a multivariate decoding analysis on all non-target object and letter stimuli in the presentation streams and examined the differences between the two task conditions. In both experiments, we found that we could decode all stimuli regardless of whether they were attended, but that attention enhanced the decodability of the relevant stimulus (object versus letter). In Experiment 1, with small letters overlaid on larger objects, attentional effects emerged around 220ms post-stimulus onset for objects, but for letters the difference started earlier, at 100ms post-stimulus onset. In a second experiment, we exchanged the position of the stimuli on the display (i.e., letters overlaid with objects) and found that the timing difference reversed accordingly. Our results show how early neural responses to simultaneously presented stimuli are modulated by certain aspects of the stimulus (e.g., size of attended stimulus) as well as our current task and attentional focus.

## 2 | METHODS

We performed two experiments that investigated the effect of attention on the representations of non-target stimuli during rapid serial visual presentation streams. Unless stated otherwise, the description of the methods below applies to both experiments. Stimuli, code, and data for this study can be found at **https://osf.io/7zhwp/**.

### 2.1 | Stimuli and design

Stimuli consisted of 16 visual objects and 16 uppercase letters (ABCDEFGJKLQRTUVY). The visual objects were coloured segmented objects obtained from www.pngimg.com spanning four categories (birds, fish, boats, and planes) with 4 images in each category. The categories could also be assigned to a two-way superordinate organisation (i.e., animals versus vehicles). The experiment was programmed in psychopy (Peirce et al., 2019). In Experiment 1, we superimposed one of 16 uppercase letters (approx. 0.8 degrees visual angle) in white font on a black circular background (Figure 1B&C) on top of the visual object stimuli (approx. 3.3 degrees visual angle). In Experiment 2, we superimposed the visual object stimuli (approx. 1.7 degrees visual angle) on one of the 16 uppercase letters (approx. 3.3 degrees visual angle) in white font on a black circular background (Figure 1D&E). Stimuli were presented in sequences of 36 (two repeats of each stimulus plus two two-back targets) for 200ms each, followed by a blank screen for 200ms. In other words, using a 2.5Hz presentation rate and a 50% duty-cycle. In alternating sequences of stimuli, participants were instructed to attend the objects or the letters and perform a cognitively demanding two-back task. Participants pressed a button whenever the stimulus they were attending to (object or letter) was the same as the stimulus that appeared two images beforehand.

We constructed 48 sequences of 32 simultaneous object and letter combinations. A sequence of stimuli was constructed by concatenating two sets of random permutations of 16 items (representing the stimuli), with the constraint that there were no repeats amongst the middle 8 items. We selected two random positions for target placement, one in the first half of each sequence and one in the second half of each sequence and inserted a target before and after the chosen positions, thus creating two-back repeats. The targets were never the same as the nearest three stimuli. Each stimulus was a target equally often. The order of stimuli in each sequence was mirror-copied, so that the order of objects and letters had matching properties while having targets in different positions. The 48 sequences were then presented twice in the experiment in random order (96 sequences in total), once for the object task, and once for the letter task. The task condition of the first sequence was counterbalanced across participants, and the conditions alternated every sequence.

A

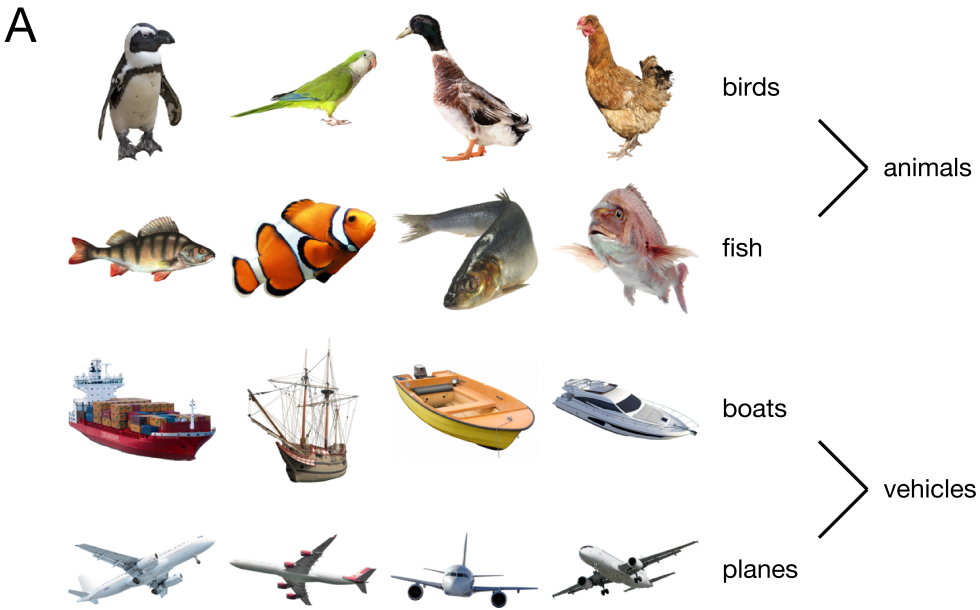

birds

animals

fish

boats

vehicles

planes

Experiment 1

B

respond to objects

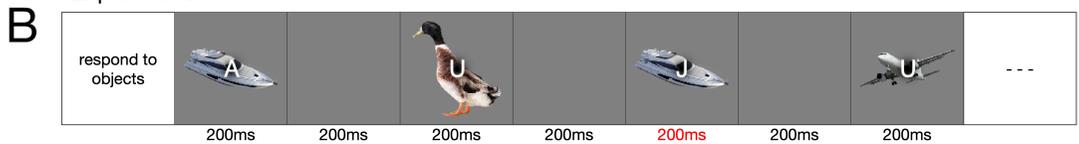

200ms 200ms 200ms 200ms 200ms 200ms 200ms

C

respond to letters

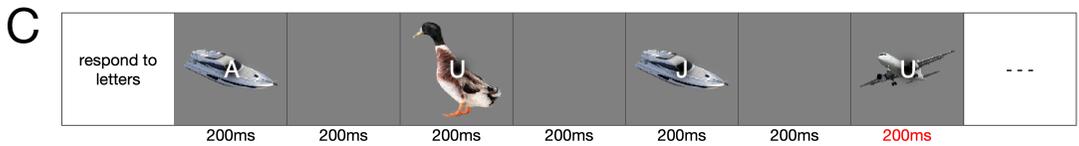

200ms 200ms 200ms 200ms 200ms 200ms 200ms

Experiment 2

D

respond to objects

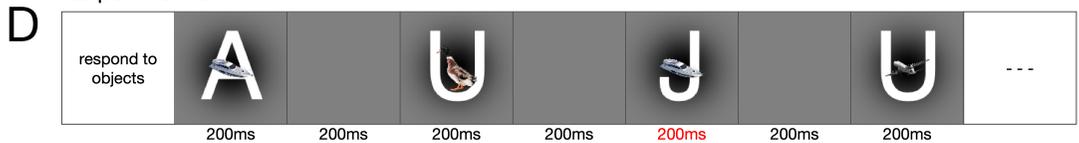

200ms 200ms 200ms 200ms 200ms 200ms 200ms

E

respond to letters

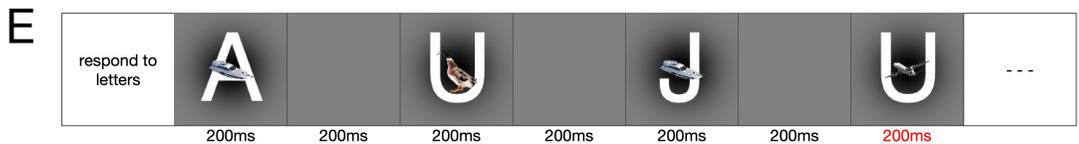

200ms 200ms 200ms 200ms 200ms 200ms 200ms

**FIGURE 1** Stimuli and design. A) Stimuli were 16 segmented objects spanning four categories (birds, fish, boats, planes) and two superordinate categories (animals and vehicles). Stimuli were presented in sequences at 2.5Hz (200ms on, 200ms off) and in each sequence, participants performed a two-back task on either the objects or on the letters. B) In the object task, participants responded with a button press when an object image was the same as the second-to-last image (two-back), while ignoring the letters. C) In the letter task, participants ignored the object images and responded on a two-back letter repeat. D, E) In the second experiment, the position of the letter and objects were swapped while keeping all other details the same.

## 2.2 | EEG recordings and preprocessing

Participants in Experiment 1 were 20 adults (9 female, 11 male; mean age 24.45 years; age range 19-41 years; all right-handed). Participants in Experiment 2 were 20 adults (17 female, 3 male; mean age 22.45 years; age range 19-36 years; 1 left-handed). All participants reported normal or corrected-to-normal vision and were recruited from the University of Sydney in return for payment or course credit. The study was approved by the University of Sydney ethics committee and informed consent was obtained from all participants. During EEG setup, participants practiced on example sequences of the two-back task. We used conductive gel to reduce impedance at each electrode side below 10 kOhm where possible. The median electrode impedance was under 10 kOhm in 35/40 subjects and under 40 kOhm in all subjects. Continuous EEG data were recorded from 64 electrodes arranged according to the international standard 10–10 system for electrode placement (Jasper, 1958; Oostenveld and Praamstra, 2001) using a BrainVision ActiChamp system, digitized at a 1000-Hz sample rate (resolution: 0.0488281μV). Scalp electrodes were referenced online to Cz. Event triggers were sent from the stimulus computer to the EEG amplifier using the parallel port. We used the same preprocessing pipeline as our earlier work that applied MVPA to rapid serial visual processing paradigms (Grootswagers et al. 2019a,b; Robinson et al. 2019). Preprocessing was performed offline using EEGlab (Delorme and Makeig, 2004). Data were filtered using a Hamming windowed FIR filter with 0.1Hz highpass and 100Hz lowpass filters, re-referenced to an average reference, and were downsampled to 250Hz. As in our previous work, no further preprocessing steps were applied (e.g., baseline correction or epoch rejection), and the channel voltages at each time point were used for the remainder of the analysis. Epochs were created for each stimulus presentation ranging from [-100 to 1000ms] relative to stimulus onset. Target epochs (task-relevant two-back events) were excluded from the analysis.

## 2.3 | Decoding analysis

To assess the representations of attended and unattended stimuli in the neural signal, we applied an MVPA decoding pipeline (Grootswagers et al., 2017) to the EEG channel voltages. The decoding analyses were implemented in CoSMoMVPA (Oosterhof et al., 2016). A regularised ($\lambda$ = 0.01) linear discriminant analysis classifier was used in combination with an exemplar-by-sequence-cross-validation approach. Decoding was performed within subject, and the subject-averaged results were analysed at the group level. This pipeline was applied to each stimulus in the sequence to investigate object representations in fast sequences under different task requirements. For all sequences, we decoded the 16 different object images, and the 16 different letters. We averaged over all pairwise decoding accuracies (i.e., bird 1 vs fish 1, bird 1 vs boat 4, bird 1 vs plane 1 etc.), such that chance-level was 50%. The analysis was performed separately for sequences from the two conditions (object task and letter task), resulting in a total of four time-varying decoding series of data per participant. For these analyses, we used a leave-one-sequence-out cross-validation scheme, where all epochs from one sequence were used as test set. We report the mean cross-validated decoding accuracies.

To determine the effect of attention on higher-level image processing, we also decoded the category (bird, fish, boat, plane) and animacy (animal versus vehicle) of the visual objects. For these categorical contrasts, we used an image-by-sequence-cross-validation scheme so that identical images were not part of both training and test set (Carlson et al., 2013; Grootswagers et al., 2019a, 2017). This was implemented by holding out one image from each category in one sequence as test data and training the classifier on the remaining images from the remaining sequences. This was repeated for all possible held-out pairs and held out sequences. The analyses were performed separately for the object and letter conditions.

## 2.4 | Exploratory channel-searchlight

We performed an exploratory channel-searchlight analysis to further investigate which features (channels) of the EEG signal were driving the classification accuracies. For each EEG channel, a local cluster was constructed by taking the closest four neighbouring channels, and the decoding analyses were performed on the signal of only these channels. The decoding accuracies were stored at the centre channel of the cluster. This resulted in a time-by-channel map of decoding for each of the contrasts, and for each subject.

## 2.5 | Statistical inference

We assessed whether stimulus information was present in the EEG signal by comparing classification accuracies to chance-level. To determine evidence for above chance decoding and evidence for differences in decoding accuracies between conditions we computed Bayes factors (Dienes, 2011; Jeffreys, 1961; Rouder et al., 2009; Wagenmakers, 2007). For the alternative hypothesis of above-chance decoding, a JZS prior was used with default scale factor 0.707 (Jeffreys, 1961; Rouder et al., 2009; Wetzels and Wagenmakers, 2012; Zellner and Siow, 1980). The prior for the null hypothesis was set at chance level. We then calculated the Bayes factor (BF), which is the probability of the data under the alternative hypothesis relative to the null hypothesis. For visualisation, we thresholded BF > 10 as substantial evidence for the alternative hypothesis, and BF < 1/3 as substantial evidence in favour of the null hypothesis (Jeffreys, 1961; Wetzels et al., 2011). In addition, we computed frequentist statistics for decoding against chance, and for testing for non-zero differences in decoding accuracies. At each time point, a Wilcoxon sign-rank test was performed for decoding accuracies against chance (one-tailed), and for the difference between conditions (two-tailed). To correct for multiple comparisons across time points, we computed FDR-adjusted p-values (Benjamini and Hochberg, 1995; Yekutieli and Benjamini, 1999).

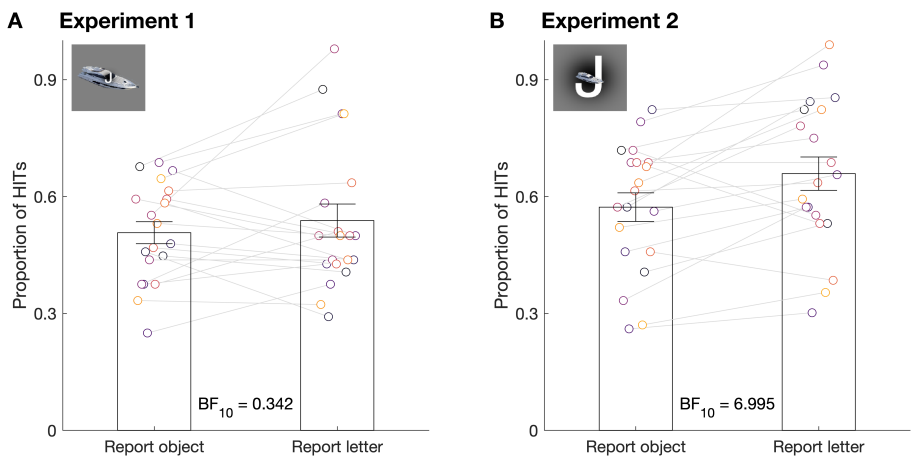

**FIGURE 2** Behavioural performance was similar between the object and letter tasks. A) Hit rate for all subjects in Experiment 1 defined as the proportion of correctly identified 2-back events. B) Hit rate for all subjects in Experiment 2. Bars show mean and standard error. Each dot represents the hit rate of one subject in one condition (object or letter task). Overall, Bayes Factors (displayed above the x-axis) indicated evidence for better performance on the letter tasks.

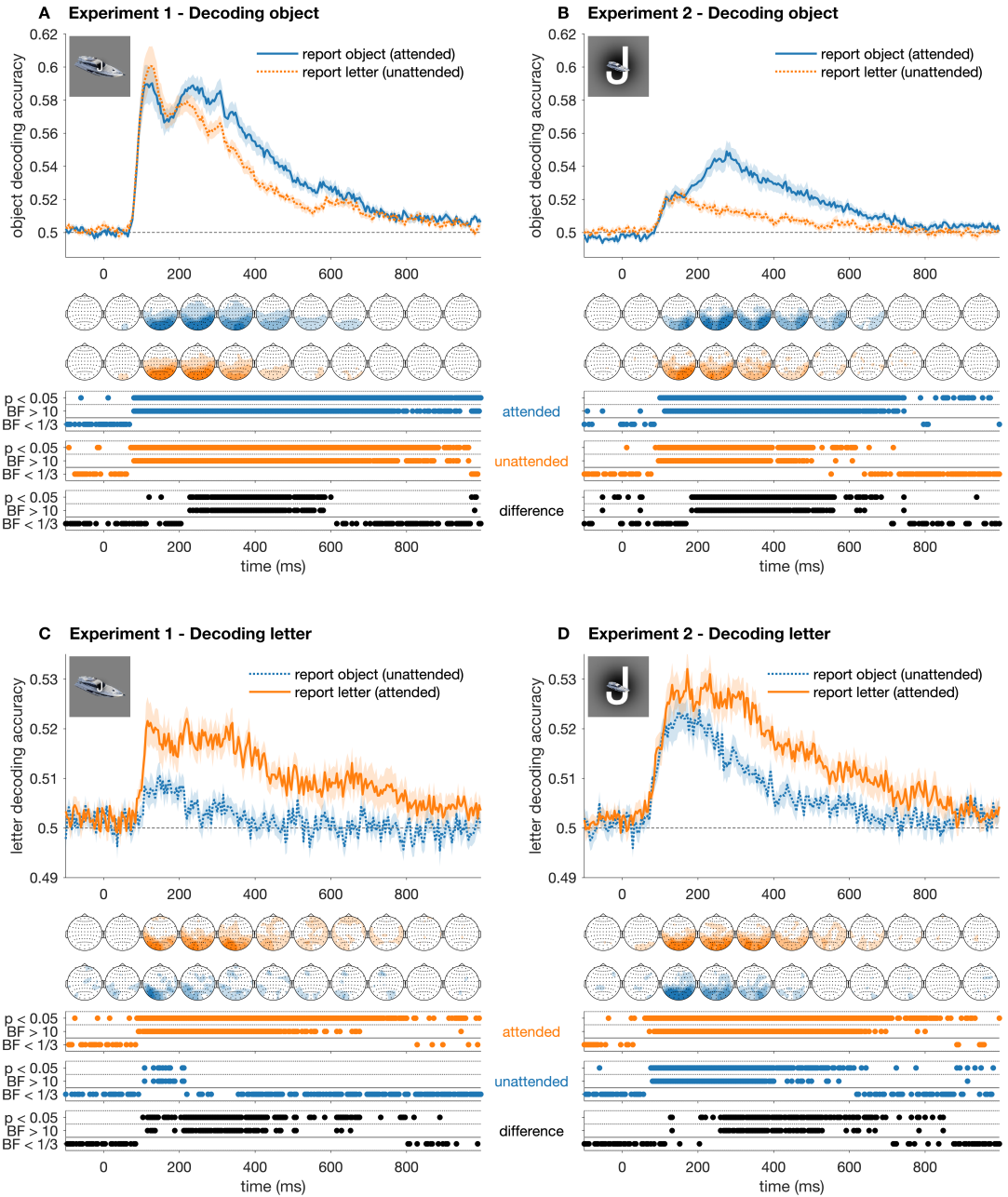

**FIGURE 3** Different effects of attention on decoding performance for objects and letters. Plots show decoding performance over time for object decoding (A&B) and letter decoding (C&D). Different lines in each plot show decoding accuracy during different tasks over time relative to stimulus onset, with shaded areas showing standard error across subjects (N = 20). Their time-varying topographies are shown below each plot, averaged across 100ms time bins. Thresholded Bayes factors (BF) and p-values for above-chance decoding or non-zero differences are displayed under each plot (note: difference lines and topographies are shown together in Figure 4). For both objects and letters, decodability was higher when they were task-relevant, but the respective time-courses of these differences varied.

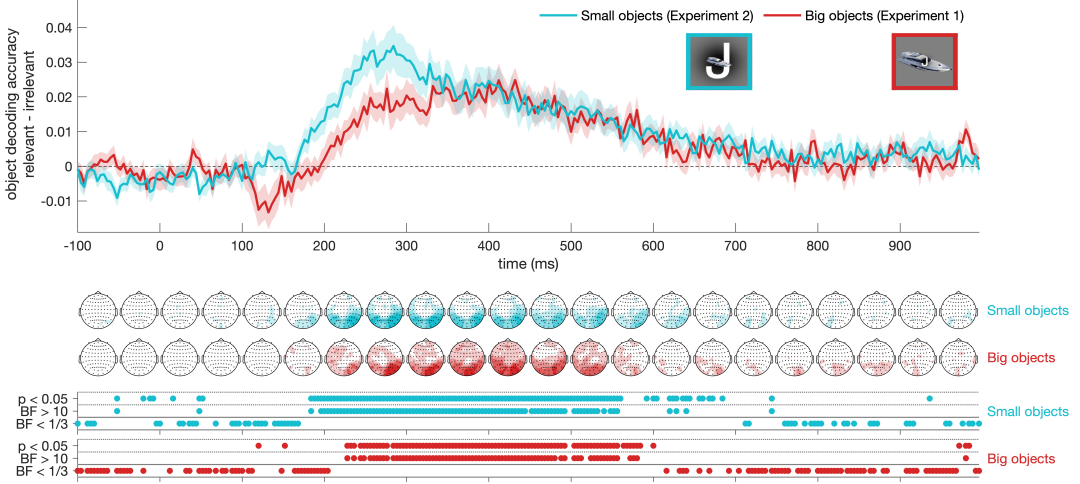

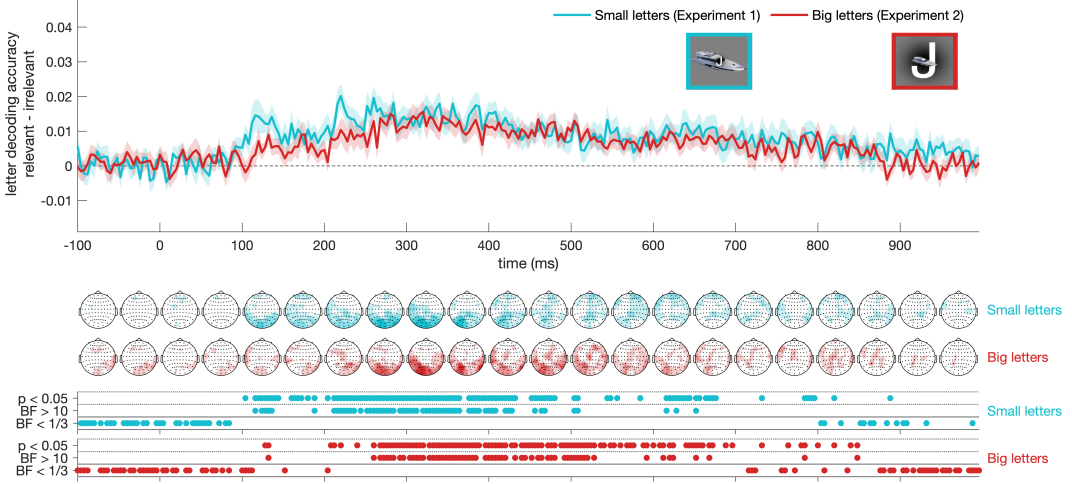

**FIGURE 4** Aggregating the attention effect over the two experiments shows the interaction between task and (relative) stimulus size. Plots shows the difference in decoding performance between task-relevant and task-irrelevant object decoding (A) and letter decoding (B). Each line reflects the mean difference from one of the two experiments relative to stimulus onset, with shaded areas showing standard error across subjects (N = 20). Their time-varying topographies are shown below each plot, averaged across 50ms time bins. Thresholded Bayes factors (BF) and p-values for above-chance decoding or non-zero differences are displayed under each plot. Note that these are the same as the stats for the non-zero difference in Figure 3. For both objects and letter stimuli, the onsets of the task-effect (relevant-irrelevant) were earlier when the stimulus was smaller.

## 3 | RESULTS

We examined the temporal dynamics of visual processing for attended (task relevant) versus unattended (task ir-relevant) stimuli that were spatially and temporally overlapping. Participants performed a difficult two-back target detection task on objects or letters simultaneously presented at fixation, and had to respond with a button press to two-back events within 5 subsequent stimulus presentations (i.e., 2 seconds) for the response to be counted as correct. Behavioural results showed that participants performed reasonably well and, on average, correctly detected 52.29% (SE 3.52) of the two-back events in Experiment 1 (Figure 2A), and 61.59% (SE 3.97) of the two-back events in Exper-iment 2 (Figure 2B). Behavioural performance was similar for detection of object (mean 50.73%, SE 2.82) and letter (mean 53.85%, SE 4.24) targets in Experiment 1 (Figure 2A) and higher for the letter (mean 65.89%, SE 4.28) than the object (mean 57.29%, SE 3.67) targets in Experiment 2 (Figure 2B). Bayes Factors indicated weak evidence for no difference in performance between task contexts in Experiment 1 (Figure 2A), and evidence for better performance on the letter task in Experiment 2 (Figure 2B).

To investigate the temporal dynamics of processing for attended and unattended stimuli, we decoded the object images and letters in every sequence, separately for the object task and letter task sequences. Figure 3 shows that ob-jects and letters were decodable regardless of whether the stimulus was attended or not, but that attention enhanced both object and letter processing. For objects, decodability was higher in the object task (task-relevant) relative to the letter task (task-irrelevant), an effect that emerged after 200ms and remained until around 600ms (Figure 3A). For letter decoding, performance was higher for the letter task than for the object task from 100ms to approximately 600ms (Figure 3C).

In Experiment 2, we exchanged the position of the object and letters on the screen, so that the letters were presented larger and overlaid with a small object at fixation. Here, attention similarly affected object and letter pro-cessing, but attention effects occurred at different times. In experiment 2, the attention effect for objects emerging after 180ms and remained until approximately 600ms (Figure 3B), and for letters occurred from 220ms to around 600ms (Figure 3D). To integrate the results from both experiments, Figure 4 shows the effect of attention for objects and letters in both experiments (i.e., the differences between decoding accuracies from Figure 3).

Combining the results from both experiments (summarised in Figure 5) shows that the attention effect started earlier for the smaller item in the display. That is, the attention effect on small letters started 100ms earlier than large letters, and the attention effect for small objects started 50ms earlier than large objects (Figure 5B). This suggests that mechanisms for attentional prioritisation are modulated by the relative retinal projection of the stimulus. The exploratory channel searchlight for object decoding (Figure 4A) suggested that the stronger decoding in the attended condition was right-lateralised. Letter decoding channel searchlights (Figure 4B) showed a more left-lateralised dif-ference in the attended condition. Together, the channel-searchlight analyses suggest that attentional effects were lateralised differently between objects and letters, but these results should be interpreted with caution as channel locations can contain information from distant sites.

To assess the effect of attention on higher-level processes, we also performed object category decoding (e.g., bird versus fish) and animacy decoding (animals versus vehicles). For both contrasts, decodable information was evident in the neural signal when objects were both attended and unattended, but information was more decodable when they were attended. Figure 6 shows that animacy and category decoding were higher for the object task compared with the letter task. Animacy (animal versus vehicle) was more decodable during the object task than the letter task from approximately 450-550ms in Experiment 1 (Figure 6A) and around 300ms in Experiment 2 (Figure 6B). In both Exper-iments, object category (e.g., bird versus fish) was more decodable from approximately 200ms (Figure 6C-D). These results show increased decoding accuracy for the more abstract categorical object information when the stimulus was relevant for the current task, but with differential effects of attention depending on the size of the stimuli.

These results are striking in showing that we can observe specific patterns of activity for the letter and object stimuli. Using a simple design, we show a clear effect of attentional enhancement on object representations that varies by the size of the images. We note that it is unlikely that the decoding results are due to artefacts (e.g., noise, eye blinks, or saccades) in the EEG signal. Firstly, for artefacts to contribute to the decoding, they would have to systematically covary with the randomly ordered 16 different stimuli over the whole experiment in every subject, which is unlikely. Secondly, stimuli were presented at very short durations at fixation, eliminating any benefit saccades would have on task performance. Finally, if the information were linked to eye movements, we would expect to see information in frontal sensors, but the channel-searchlight showed information most prominently originated in occipital sensors.

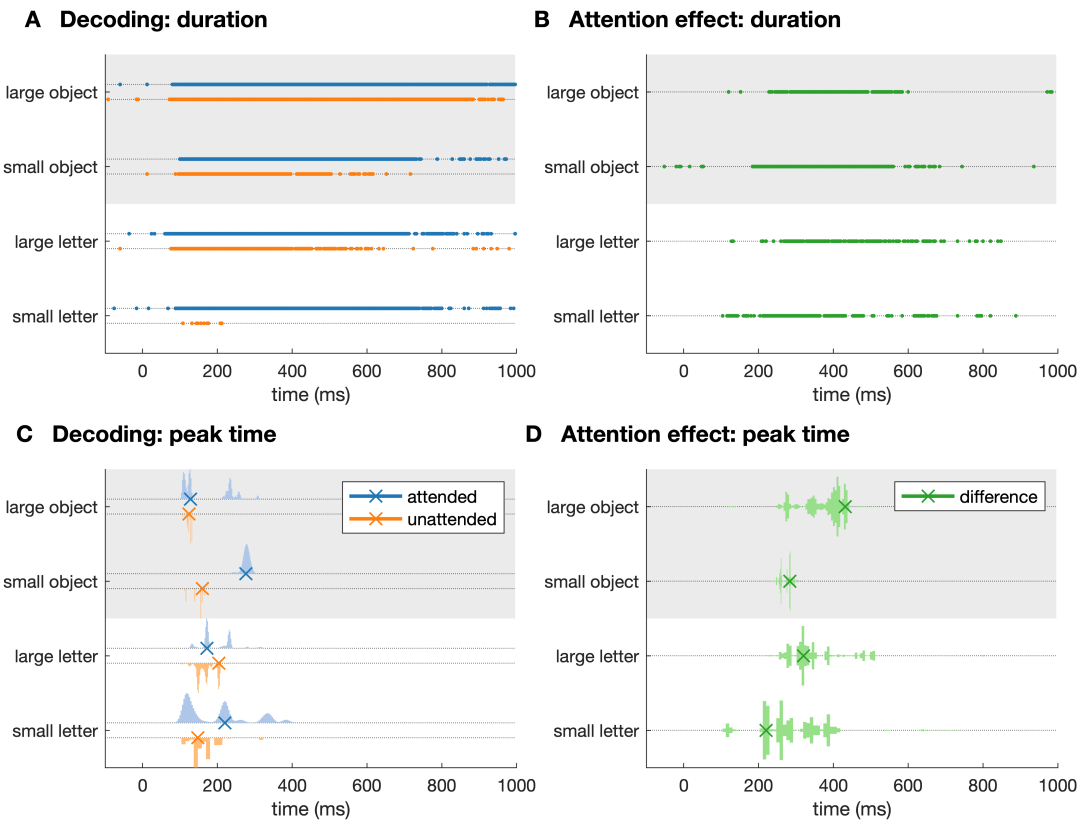

**FIGURE 5**   Summary of main findings. The top row (A,B) shows the significant time points for each contrast. The bottom row (C,D) shows the time of the peak (denoted by x) accompanied by the distribution of peak times obtained by drawing 10,000 samplings from the subjects with replacement. Left columns (A,C) show results for decoding against chance, and right columns (B,D) show the difference between attended and unattended decoding.

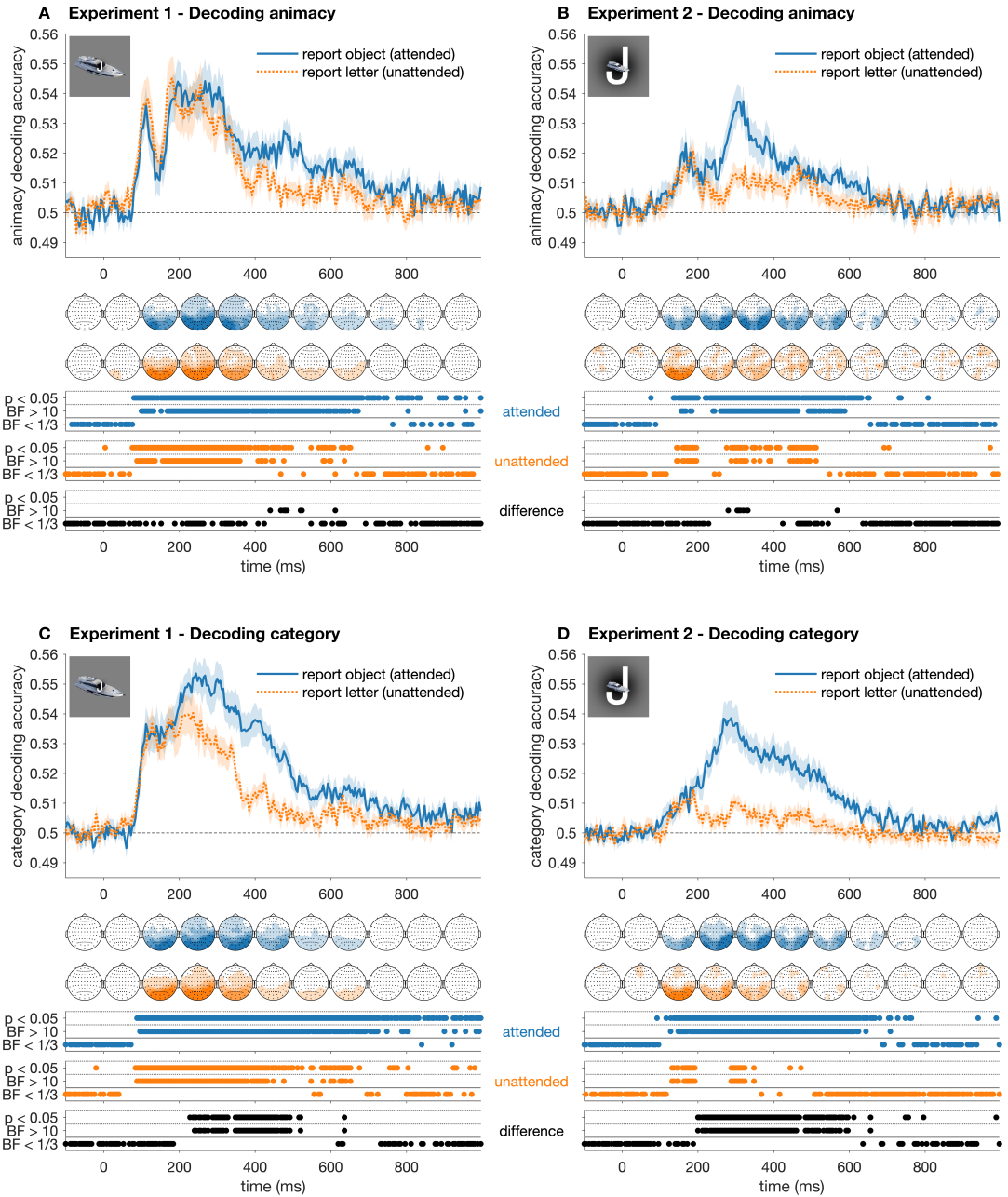

**FIGURE 6** Effect of attention on higher level categorical object contrasts in Experiment 1 were similar to individual object decoding. Plots show decoding performance over time for object animacy decoding (A) and object category decoding (B). Different lines in each plot show decoding accuracy during different tasks over time relative to stimulus onset, with shaded areas showing standard error across subjects (N = 20). Their time-varying topographies are shown below each plot, averaged across 100ms time bins. Thresholded Bayes factors (BF) and p-values for above-chance decoding or non-zero differences are displayed under each plot.

## 4 | DISCUSSION

In this study, we asked how attention modulates the representations of visual stimuli. Participants monitored streams of letters overlaid on objects (Experiment 1) or objects overlaid on letters (Experiment 2) and performed a 2-back target detection task on either the letters or the objects. Importantly, we did not analyse the responses to the 2-back targets, but rather investigated how task context influenced the representation of all other stimuli in the streams. We predicted that attended and relevant information would be strongly affected by attentional prioritisation processes in the difficult 2-back task. Remarkably, we could decode all attended and unattended stimuli in both experiments, even though they were spatially and temporally overlapping, but the dynamics of the representations varied according to the task and the size of the stimuli. As predicted, we found in general that attending to objects improved the decoding accuracy for objects and that attending to letters improved the letter decoding accuracy. However, the time course of these attentional effects varied, such that the improved decoding of task-relevant information emerged after 200ms for large stimuli, but before 200ms for small stimuli (Figure 5). Taken together, these findings show that task context selectively enhances the processing of relevant visual stimuli, and that this effect is specific to the features of the stimuli being selected.

Our results shed light on the time course of attention and on the ongoing debate in the literature about which stages of processing are affected by attentional mechanisms (cf. Alilović et al., 2019; Baumgartner et al., 2018). The observation that size affects the temporal dynamics of attention on early representations is consistent with previous studies showing early effects when varying the spatial aspects of the task (Mangun, 1995; Wyart et al., 2015). In contrast, previous work has used similar 2-back tasks on letters overlaid on images to divert attention away from the images and found that the early processing of the images was not affected (e.g., Groen et al., 2015). However, this work did not analyse the responses to the letters (i.e., the small items in their display), which our study showed to be affected early in the response. The later and more prolonged effects of attention on the decoding accuracies is consistent with work showing enhancements of target versus distractor (i.e., task-relevant stimulus information) coding starting around 200ms (Kranczioch et al., 2005; Marti and Dehaene, 2017). These later (>200ms) sustained effects are consistent with work suggesting information processing up to 200ms is mainly driven by feedforward mechanisms, and that after around 200ms, a wider network of recurrent processes is recruited to generate behavioural decisions (Dehaene and Changeux, 2011; Gwilliams and King, 2020; Lamme and Roelfsema, 2000). Our results complement these findings by showing how attentional effects interact with visual features of the stimulus (relative size), which highlights how attention can impact representations at different information processing stages. Future work can build on our findings by further investigating the interactions between early and late attentional effects.

All stimuli in this study evoked distinct patterns of neural responses regardless of whether they were relevant to the task at hand. That is, letters and objects were decodable in all conditions. This fits with our previous work showing that task-irrelevant objects can be decoded from rapid streams (Grootswagers et al., 2019a; Robinson et al., 2019), likely reflecting a degree of automaticity in visual processing and confirming that target selection is not a requirement for stimulus processing. The current study extends these findings by showing that two simultaneously presented visual objects are decodable even when one stimulus is much less prioritised than the other due to task demands and stimulus size. Strikingly, the duration of above chance decoding was much longer than the stimulus presentation time. The long, sustained decoding of attended stimuli could reflect the requirement of holding stimuli in memory for two subsequent presentations (i.e., >800ms) to perform the 2-back task. An easier task with a smaller working memory component may result in different attentional effects, for example it may reduce the need for suppression of the irrelevant stimulus. However, the memory component of the task cannot fully account for the prolonged effects, as for objects, sustained above chance decoding was observed even when the stimulus was not attended, an observation that is consistent with our previous work on information coding rapid sequences (Grootswagers et al. 2019a,b; King and Wyart 2019; Robinson et al. 2019). For example, unattended object information was above chance for up to 900ms post stimulus-onset in Experiment 1 (Figure 3A), and up to 600ms in Experiment 2, when the objects were smaller (Figure 3B). This shows that visual information was maintained in the system even though it was not task relevant and it was presented in conjunction with a task-relevant stimulus. Thus, task-irrelevant information appeared to reach higher levels of processing than just feature processing, even though it was not the subject of attention. Indeed, category and animacy decoding (Figure 6) suggests that object stimuli were processed up to abstract levels in the visual hierarchy. In sum, all objects and letters were decodable even during fast-changing visual input and even when they were not attended. Importantly, however, we found that attention enhanced the distinctiveness (i.e., decodability) of the attended visual stimuli.

Attention affected both the strength and duration of evoked visual representations. For both letters and objects, decodability was higher and prolonged when they were task-relevant compared to when they were irrelevant. This is particularly striking because the letter and object tasks involved exactly the same sequences of images and analyses, so differences in decoding arise exclusively from the attentional focus imposed by the task that participants performed. Furthermore, it is important to note that target images (i.e., the two repeating stimuli) were not analysed, meaning that target selection and response processes were not contained within our results. The difference we observed thus can mainly be attributed to attentional mechanisms. The enhancement of attended object information around 220ms is consistent with evidence of effects from the attentional blink and target selection literature, which has often reported differences in N2 and P300 ERP components (Kranczioch et al. 2007, 2003; Sergent et al. 2005). Target stimuli in rapid streams have been found to evoke stronger signals around 220ms (Marti and Dehaene 2017). In these designs, however, it is difficult to distinguish between the effects of target-selection and the enhancement of task-relevant information. As all our analyses were performed on non-target stimuli, our results point towards a general enhancement of task-relevant stimuli at this time scale, even for images that are not selected for target-processing. This points towards a more general enhancement effect of task-relevant information occurring around 220ms that supports flexible task performance in many paradigms.

Attentional effects on the letter stimuli followed a different trajectory to that of the objects, with an onset around 100ms for letters versus 220ms for objects in Experiment 1. This could be explained by the letters comprising a smaller part of the stimulus arrangement. Previous work has shown effects of eccentricity on neural responses (e.g., Eimer 2000; Isik et al. 2014; Müller and Hübner 2002), but our results could also be attributed to differences in spatial attention allocated to the letter versus image task. Indeed, when we exchanged the stimulus position in Experiment 2, we observed an earlier onset of the attentional effects on object decoding, but the effect for letters seemed to occur later. Channel searchlight analyses further suggested that the attentional effects were more left lateralised for the letter task, and right lateralised for the object task. The regions of highest decoding in the searchlights do not necessarily reflect the regions where the signal originates, but these results do fit with previous work showing that letter processing is typically left lateralised (Cohen et al. 2003; Puce et al. 1996), and that animate objects tend to evoke right hemisphere dominant responses (Bentin et al. 1996; Puce et al., 1996, 1995). The different spatio-temporal dynamics between the enhanced decodability of relevant information between the object and letter tasks suggest that attentional effects are dependent on perceptual characteristics of the specific stimuli being processed.

For objects and their conceptual category decoding, we found evidence for no attentional effect on the initial responses (until around 180ms). This is consistent with recent work that reported no evidence for attentional effects on early visual ERP components or decoding accuracies (Alilović et al. 2019; Baumgartner et al. 2018). In contrast, we did find attentional effects on decoding accuracy for the earliest responses to letters (Figure 3C), which were more decodable throughout the epochs when task relevant. One explanation of this difference is that objects are auto-matically and unconsciously processed, but letters may require an active recruitment of their respective processing mechanisms. Alternatively, the object stimuli used here are visually much more distinct (different colours and shapes) than the letter stimuli which facilitates decoding of visual feature differences.

In conclusion, we found that attention enhances the representations of task-relevant visual stimuli, even when they were spatially and temporally overlapping with task-irrelevant stimuli, and even when the stimuli were not selected as target. Our results suggest that attentional effects operate on the specific perceptual processing mechanisms of the stimulus, differing across stimulus type and size. This points towards a multi-stage implementation of information prioritisation that guides early perceptual processes, as well as later-stage mechanisms.

## Acknowledgements

This research was supported by ARC DP160101300 (TAC), ARC DP200101787 (TAC), and ARC DE200101159 (AKR). The authors acknowledge the University of Sydney HPC service for providing High Performance Computing resources.

## Conflict of interest

The authors declare no conflicts of interests.